\documentclass{aa}
\usepackage[varg]{txfonts}
\usepackage{amsmath}
\usepackage{amssymb}
\usepackage{graphicx}% Include figure files
\usepackage{dcolumn}% Align table columns on decimal point
\usepackage{bm}% bold math
\usepackage{ulem}

\begin{document}

\title{On the crystallization onset in white dwarfs}
\author{D. A. Baiko\thanks{E-mail:baiko.astro@mail.ioffe.ru}}
\institute{Ioffe Institute, Politekhnicheskaya 26, 194021 Saint Petersburg, Russia}
\date{Received/accepted}
%\pagerange{\pageref{firstpage}--\pageref{lastpage}} \pubyear{2014}
%\maketitle
%\label{firstpage}
%
%\begin{abstract}
%
\abstract{Thermal evolution of the central region of a $0.9 \, M_\odot$ C/O white 
dwarf at the initial stage of the ion mixture crystallization is studied 
by numerically solving the heat equation on a fine spatial and temporal 
grid and by including a detailed treatment of the latent heat release.
Formation of two spherical shells is observed. The outer one surrounds 
a region where crystallization has begun. The inner one bounds a fully 
solidified core which has exhausted its latent heat. The region between
the shells is partially liquid and partially solid. It gradually emits 
the latent heat of crystallization and also it releases light elements 
(carbon) 
in the process of element redistribution, accompanying the mixture 
solidification. Assuming that all released light elements cross the 
outer shell, we have estimated their flux induced by the mixture 
crystallization. The resulting flux is not divergent and is much smaller 
than an estimate derived from the growth rate of the fully crystallized 
core.}
%
%\end{abstract}

\titlerunning{On the crystallization onset in white dwarfs}
\authorrunning{Baiko}

\maketitle
%\voffset=-0.8in
%%%%%%%%%%%%%%%%%%%%%%%%%%%%%%%%%%%%%%%%%%%%%%%%%%%%%%%%%%%%%%%%%%%%

\section{Introduction}
The origin of strong magnetic fields in white dwarf stars (WD) is a hot
topic of modern astrophysics \citep[e.g.][]{BL22}. In particular, a 
debate is ongoing on whether the magnetic field can be generated in a WD 
core via crystallization driven convection and dynamo and whether it can 
subsequently emerge and be observed at the stellar surface 
\citep[e.g.][and references therein]{I+17,G+22,F+23,MD24,BG24}. Most 
recently, \citet*{FCC24} have proposed a short-lived ($\lesssim 10$ Myr) 
phase of efficient compositionally driven convection characterized by a 
large upward light element flux and rapid fluid motions, operating at 
the initial stage of C/O WD crystallization. \citet{L+24} discussed a 
powerful dynamo driven by rapid upward flow of $^{22}$Ne-depleted solid 
flakes in the process of $^{22}$Ne distillation prior to core 
crystallization in C/O/$^{22}$Ne WD.  

The present paper focuses on crystallizing C/O WD. We study the thermal 
evolution of matter in the inner core region at the freezing onset by 
numerically solving the heat equation on a fine spatial and temporal 
grid and by including a detailed treatment of the latent heat release.
As a by-product, we obtain an alternative estimate of the light element
flux induced by the C/O mixture crystallization.

\section{Model}
\label{model}
When freezing of a WD core begins, the very central region reaches 
its crystallization temperature and then it cannot cool 
further\footnote{This statement is obvious for dense matter composed of a single ion sort.
For an ionic mixture, it is true, if the ion composition of the freezing 
liquid remains fixed. In what follows, we shall make this assumption
as it is inherent in the effect to be considered (see below).} 
unless all of its latent heat has been released and removed.
The release of the 
latent heat is controlled by the ability of the system to carry it away, 
that is by available thermal gradients and heat fluxes. As the time goes 
on, more outward layers also reach their crystallization temperature and 
start releasing latent heat i.e. the region that has begun 
crystallization gradually expands. At some point, the central part 
completely exhausts its latent heat and resumes cooling according to the 
standard heat equation. Thus, in general, once crystallization in a WD 
begins, there are three regions. In the inner and outer regions, matter 
cools according to the standard heat equation being, respectively, in a 
crystal or liquid state. The intermediate region represents a reservoir 
of heat, where matter is partially solid and partially liquid, its 
temperature being equal to the local crystallization 
temperature\footnote{This is similar to a water-ice mixture at 
0$^\circ$C.}. To model the intermediate region, one simply has to keep 
track of the balance between the total amount of available heat and 
incoming and outgoing heat fluxes. 

Since we are mostly interested in a qualitative picture of the freezing 
onset, our model is significantly simplified. Following \citet{FCC24}, 
we assume a specific form of the mass density $\rho$ dependence on the 
radial coordinate $r$:
\begin{equation}
     \rho = \rho_{\rm c} \left(1-\frac{r^2}{R^2_{\rm WD}}\right)^{3/2}~,
\end{equation}
where $\rho_{\rm c} = 1.86 \times 10^7$ g/cc is the central 
density and $R_{\rm WD} = 6371$ km is the WD radius. The specific 
values are taken from the evolutionary model\footnote{
\url{https://www.astro.umontreal.ca/~bergeron/CoolingModels}} of a 
WD with a thick 
hydrogen envelope and a mass of $M_{\rm WD} = 0.9 \, M_\odot$ near the 
crystallization onset \citep[][the reference model]{B+20}. 

We further assume equal proportions by mass of C and O but neglect any effects of phase 
separation and element redistribution, including the gravitational 
energy release, on thermal evolution. 
In reality, the liquid around newly forming crystals gets depleted of 
the heavier element (oxygen) and enriched with the lighter element 
(carbon). This creates 
%convective 
compositionally driven
instability conditions discussed e.g. 
by \citet{F+23,MD24}, resulting in the light 
element flux $F_X$ directed upward and an opposite heavy 
element flux, which replenishes the oxygen content of the liquid. Since 
we limit our consideration to the inner core region and to initial 
stages of crystallization, we do not describe these effects 
in detail except for estimating the flux $F_X$ and adopting the fixed liquid
composition assumption.    
We expect that these effects, 
particularly the extra heat, produced by mixing, which also has to be 
removed, 
could only strengthen our conclusions. 
Furthermore, we neglect any effects 
associated with distillation of $^{22}$Ne \citep*[][]{I+91,BDS21} or any 
other heavier species.  

The crystallization (melting) temperature $T_{\rm m}$ is calculated as  
\begin{equation}
          T_{\rm m} = \frac{Z_1^{5/3} e^2}{a_{\rm e} k_{\rm B} 
          \Gamma_{\rm 1m}} \xi_{\rm CO}(3/7) ~,
\end{equation}
where $Z_1=6$ is the carbon charge number, 
$a_{\rm e} = (4 \pi n_{\rm e}/3)^{-1/3}$, $n_{\rm e}$ is the electron 
density, $\Gamma_{\rm 1m} =175.6$ is the Coulomb coupling parameter of 
a classic one-component plasma at melting \citep[][]{BC22}, and 
$\xi_{\rm CO}(3/7)=1.05$ is a correction at the oxygen number fraction 
$x_{\rm O} = 3/7$ based on the phase diagram of the C/O mixture 
\citep[][]{BD21}. 

Furthermore, we assume a constant heat capacity $C=2.7 k_{\rm B}$ per 
ion, which is a reasonable approximation for mildly quantum liquids and 
solids (in our system, ion plasma temperature is $\sim 3$ 
times greater\footnote{The ion plasma temperature  
$T_{\rm p}\equiv\hbar \omega_{\rm p}/k_{\rm B} \approx 2.2 \theta_{\rm D}$, 
where $\omega_{\rm p}$ is the ion plasma frequency and $\theta_{\rm D}$
is the Debye temperature. At $T_{\rm p}/T \gg 1$, one-component ion 
plasma is known to be in the regime of strong ion quantum effects.} 
than the actual temperature) in the vicinity of the phase transition 
\citep[cf.][]{BC22}; constant thermal conductivity 
$\kappa=1.2 \times 10^{16}$ erg cm$^{-1}$ 
s$^{-1}$ K$^{-1}$ based on the electron-ion collision frequency 
calculation, fit, and extrapolation to mixtures \citep[][]{PBHY99}; and 
the latent heat of crystallization for the classic equimass C/O mixture, 
$\lambda=0.67 \,k_{\rm B} T_{\rm m}$ per ion \citep[][]{B23}.

Realistic study of the crystallization onset implies a description of 
the local latent heat release and removal which requires a 
non-isothermal approach. In our model, the initial temperature at the 
stellar center is taken to be 0.03\% greater than the melting 
temperature $T_{\rm m}(0)$. The initial temperature profile, 
$T_{\rm ini}(r)$, is based on the following scaling:
\begin{eqnarray}
     T_{\rm ini}(r) &=& 1.0003 \, T_{\rm m}(0) - 0.5 K r^2~,
\nonumber \\
     K &=& \frac{\rho \, C}{3 \kappa \langle A \rangle m_{\rm u}}
               \left|\frac{{\rm d}T}{{\rm d}t}\right|~,       
\label{Kdef}
\end{eqnarray}
where $\langle A \rangle$ is the average mass number of the mixture,
$m_{\rm u}$ is the atomic mass unit, and ${\rm d}T/{\rm d}t$ is a
time derivative of temperature. 

We analyze thermal evolution of matter 
only out to $r=r_0$ with $r_0=3000$ km to avoid the need to describe 
complicated physics of the outer, partially ionized and partially 
degenerate layers. Consequently, a boundary condition at $r_0$ is 
required and for that we specify the time derivative of temperature, 
${\rm d} T (r_0)/{\rm d} t = \alpha_0(t)$. In principle, the true form 
of $\alpha_0(t)$ can be obtained only from a full-star evolutionary 
modeling with a high-level code and with account of the effects 
investigated in the present paper. In the absence of that, we can, 
first, assert that $T(r_0)$ monotonously decreases with $t$ and, second, 
take advantage of the fact that we are actually interested in a 
relatively brief time interval $\Delta t < 100$ Myr since the 
crystallization onset, which covers the purported efficient convection 
stage. For this reason, the time derivative is assumed to be constant 
$\alpha_0=-1.5 \times 10^{-10}$ K/s. This value of $\alpha_0$ is close 
to the average time derivative of the central temperature in the 
reference model shortly {\it before} the crystallization onset. On the 
other hand, in our model, such a value of $\alpha_0$ results in the 
average time derivative of the central temperature {\it after} the 
crystallization onset equal to $-0.83 \times 10^{-10}$ K/s, which is 
practically the same as the one in the reference model. Besides that,
we have tried several other values of $\alpha_0$, bracketing 
the fiducial value above within a factor of $\sim 2$, which, naturally, 
yielded qualitatively similar picture of the temperature variation. We 
therefore do not expect that implementing a weak decrease of 
$|\alpha_0(t)|$ over considered $\Delta t$ could affect our main 
conclusions. We have also set ${\rm d}T/{\rm d}t$ to $\alpha_0$ in 
equation (\ref{Kdef}).  

%%%%%%%%%%%%%%%%%%%%%%%%%%%%%%%%%%%%%%%%%%%%%%%%%%%%%%%%%%%%
\begin{figure}
%\begin{figure*}
%\centering
\begin{center}
%\leavevmode
%\includegraphics[bb=71 508 569 741, width=\textwidth]{assemble.ps}
%\includegraphics[bb=11 1 736 557, width=84mm]{Tevol_7CO.eps}
%\includegraphics[bb=22 1 736 728, width=84mm]{Tevol_7CO.eps}
\includegraphics[bb=44 0 735 1308, width=88mm]{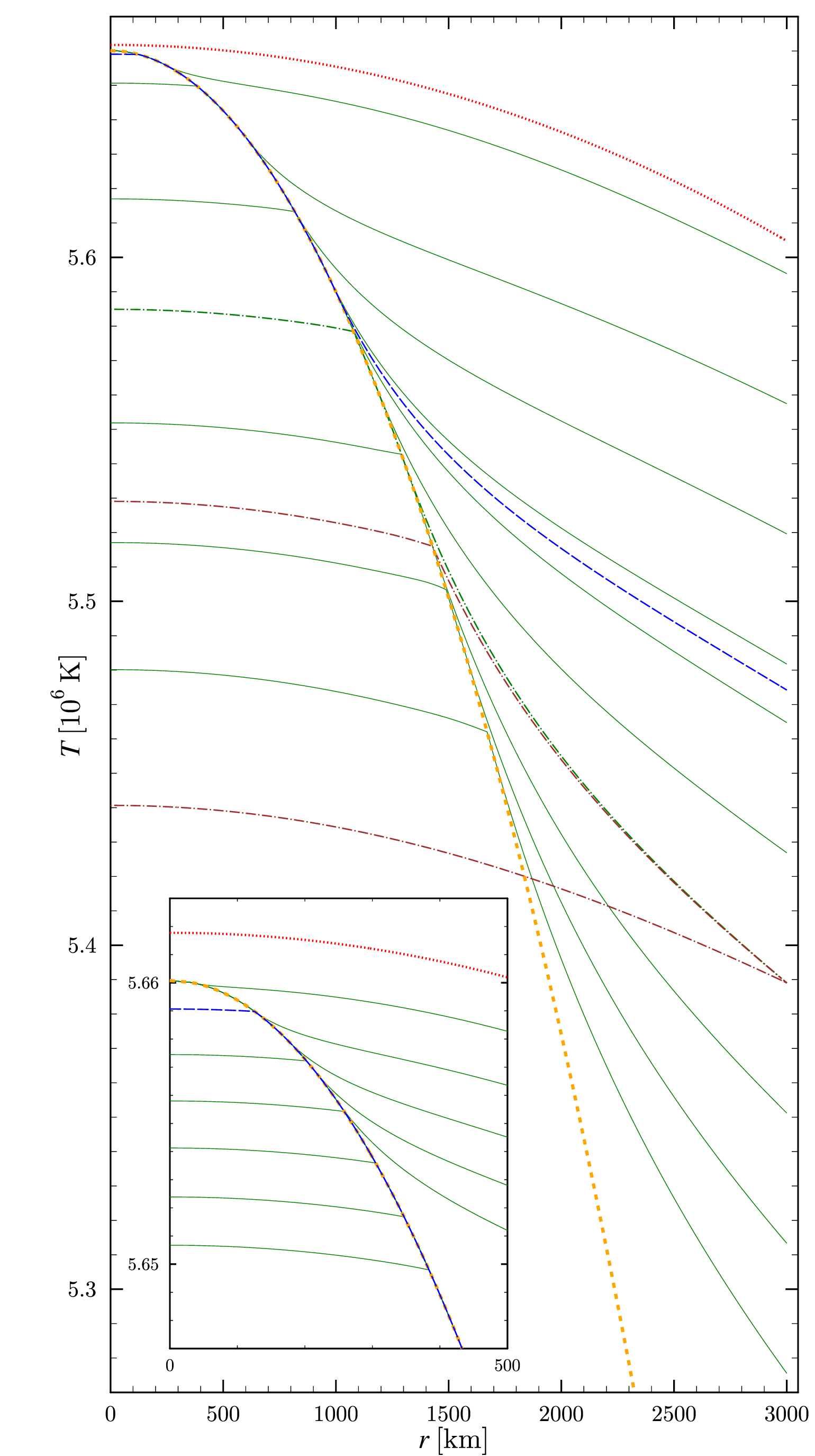}
\end{center}
\vspace{-0.2cm}
\caption[ ]{Internal temperature profile of cooling WD model at 
various times. Initial profile (dotted red), profile shortly after
formation of the fully crystallized core (long dashed blue), three 
profiles at 45.6 Myr since the beginning of modeling (dot-dashed green 
for the default latent heat value, dot-dashed brown for two times smaller 
and zero latent heat values), profiles at other times (solid green, see 
text for details). Crystallization temperature is shown by thick short  
(orange) dashes.} 
\label{profile}
%\end{figure*}
\end{figure}
%%%%%%%%%%%%%%%%%%%%%%%%%%%%%%%%%%%%%%%%%%%%%%%%%%%%%%%%%%%%

\section{Results}
The result of modeling is shown in Fig.\ \ref{profile}. Displayed is 
the temperature profile as a function of $r$ with different curves 
corresponding to different time stamps. The dotted (red) curve 
demonstrates the initial profile. Four solid (green) curves between it and the 
long dashed (blue) profile show the evolution before the latent heat in the 
center is exhausted: the upper solid curve is separated by 2 Myr from the
initial one, the other three curves add 8 Myr of evolution each. 
On the left-hand side, these solid curves approach and then smoothly merge 
onto the thick short dashed (orange) line, which is the profile of 
the crystallization temperature. At $r=0$, they take the value $T_{\rm m}(0)$. 
Hence, in this case, there are only two regions: the inner one is at the local 
crystallization temperature and the outer one represents the cooling 
liquid. In the inset, the top-left corner of the graph is shown in more 
detail. Five (green) solid profiles to the right of the crystallization 
temperature are separated from each other and from the 
initial (dotted red) one by 400 kyr.

The long dashed (blue) curve corresponds to 27.6 Myr since
the beginning of the simulation, slightly later than the moment at 
which the crystallization of the innermost core region is complete i.e. 
all its latent heat is removed.
Consequently, the solid (green) curves below it contain three segments,
corresponding to the three regions described in the first paragraph of 
section \ref{model}. The segments, describing the cooling crystal with progressively
lower central temperatures, can be seen to the left of the crystallization profile.
The mutual time separation between these curves is 8 Myr (the 
dot-dashed green curve is also part of this family, see below) and the upper 
one of them is at 2 Myr after the long dashed profile. This regime is 
characterized by an expansion of the region which has reached 
$T_{\rm m} (r)$ and of the region which has exhausted its latent heat
and cooled below $T_{\rm m} (r)$. The radial extent of the intermediate
region shrinks and by the last curve, which displays the temperature 
profile at 69.6 Myr, its size becomes smaller than the spatial grid. 
Five solid (green) curves to the left of the crystallization
temperature profile in the inset are separated from each 
other and from the long dashed (blue) curve by 400 kyr. 

For illustration, by dot-dashed curves we show the temperature profile 
at 45.6 Myr for three sequences with 
different latent heat values (assuming the same initial profile and 
boundary condition): the upper one (green) corresponds to the standard 
$\lambda$ value, the one in the middle and the lower one (both brown)
demonstrate the cases with $\lambda=0.33 \,k_{\rm B} T_{\rm m}$ and zero 
latent heat, respectively. For the latter, there should be no difference 
between the liquid and solid phases in our code, and a smooth curve,
cooling rapidly in the center, is in fact obtained.  

\begin{figure}
%\begin{figure*}
\begin{center}
%\leavevmode
%\includegraphics[bb=71 508 569 741, width=\textwidth]{assemble.ps}
\includegraphics[bb=14 0 728 564, width=84mm]{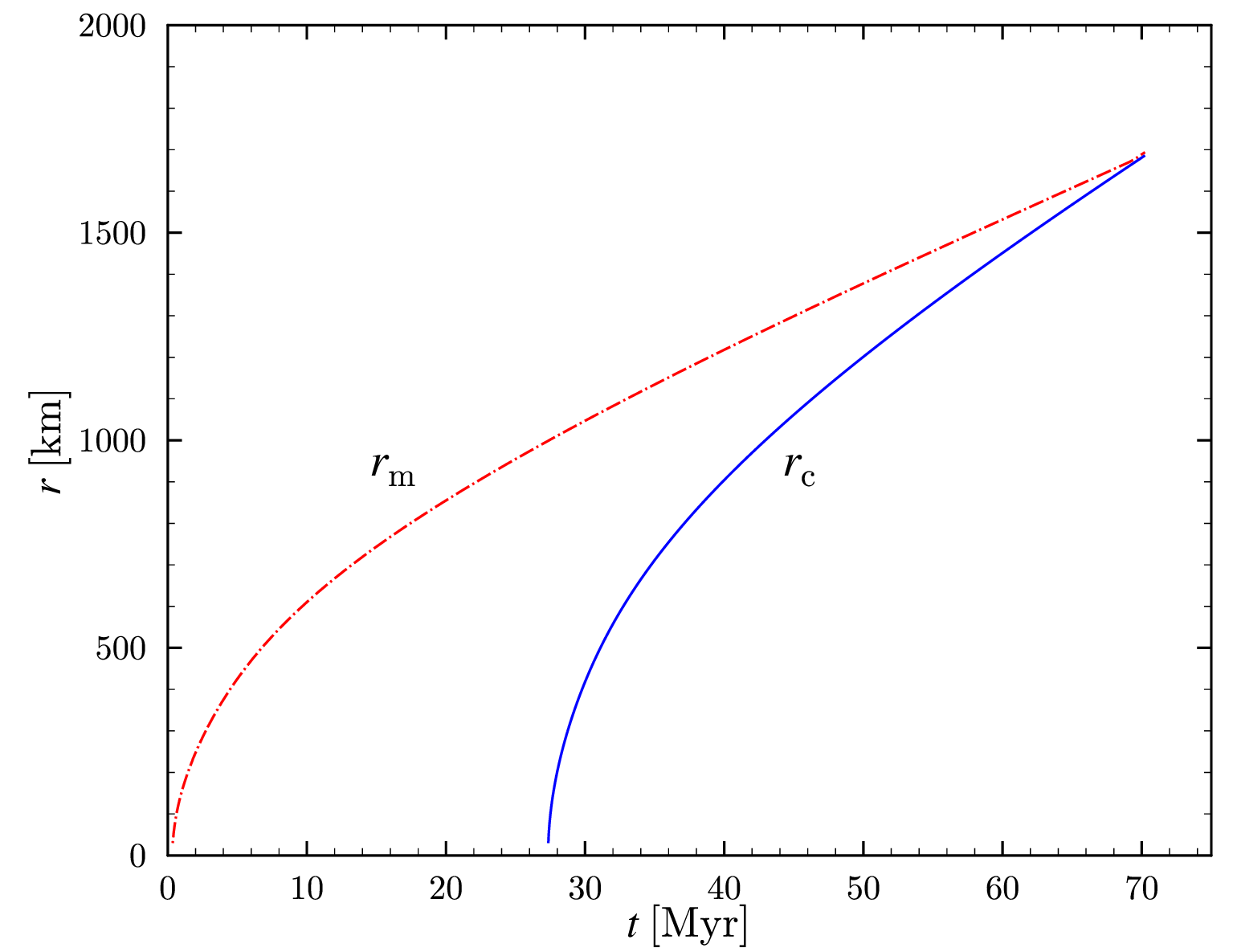}
\end{center}
\vspace{-0.2cm}
\caption[ ]{Radius of the region, which has started crystallization 
(dot-dashed red), and of the fully crystallized core (solid blue) vs. 
time since the beginning of modeling.} 
\label{radius}
%\end{figure*}
\end{figure}

In Fig.\ \ref{radius}, we show radii of the regions that have started
and completed crystallization denoted as $r_{\rm m}$ and $r_{\rm c}$,
respectively, by dot-dashed (red) and solid (blue) 
curves as functions of time. As we observed in Fig.\ \ref{profile}, the 
difference between the initial time stamps for the 
two processes is $\sim 27$ Myr. At the beginning, both dependences have 
infinite time derivative in a qualitative agreement with the analytic 
model of \citet{FCC24}. Eventually, the two curves merge (to within the spatial 
grid size).    

%%%%%%%%%%%%%%%%%%%%%%%%%%%%%%%%%%%%%%%%%%%%%%%%%%%%%%%%%%%%
%%%%%%%%%%%%%%%%%%%%%%%%%%%%%%%%%%%%%%%%%%%%%%%%%%%%%%%%%%%%

Let us now estimate the flux of light elements, $F_X$, resulting from phase
separation accompanying crystallization (in our case, this is the flux of 
carbon ions). \citet{FCC24} linked the flux with the 
velocity of the crystallization front. Specifically, if the front moves 
from $r-{\rm d}r$ to $r$ over a time interval ${\rm d}t$, then, in 
their model, the entire spherical shell of thickness ${\rm d}r$ turns 
from the liquid to the solid phase. This allows one to calculate the 
mass of released light elements, which then cross the sphere of the 
radius $r$ 
during ${\rm d}t$ (by convective motions). This quantity was found to be 
divergent at the initial stages of crystallization. 

However, as our solution shows, such a picture does not apply at the 
crystallization onset. Actually, light elements are released rather 
slowly 
within the spherical shell of the outer radius $r_{\rm m}$ and inner 
radius 
$r_{\rm c}$ ($r_{\rm c}=0$ for the first $\sim 27$ Myr). The mass of 
released light elements over an interval ${\rm d}t$ is proportional to 
the total mass which has solidified. This, in turn, is proportional to 
the amount of latent heat carried away from this region. If convection 
is sufficiently rapid, then the light elements can cross the sphere at 
$r_{\rm m}$ during ${\rm d}t$. This provides an upper estimate of the 
convective flux.

\begin{figure}
%\begin{figure*}
\begin{center}
%\leavevmode
%\includegraphics[bb=71 508 569 741, width=\textwidth]{assemble.ps}
\includegraphics[bb=11 0 728 564, width=84mm]{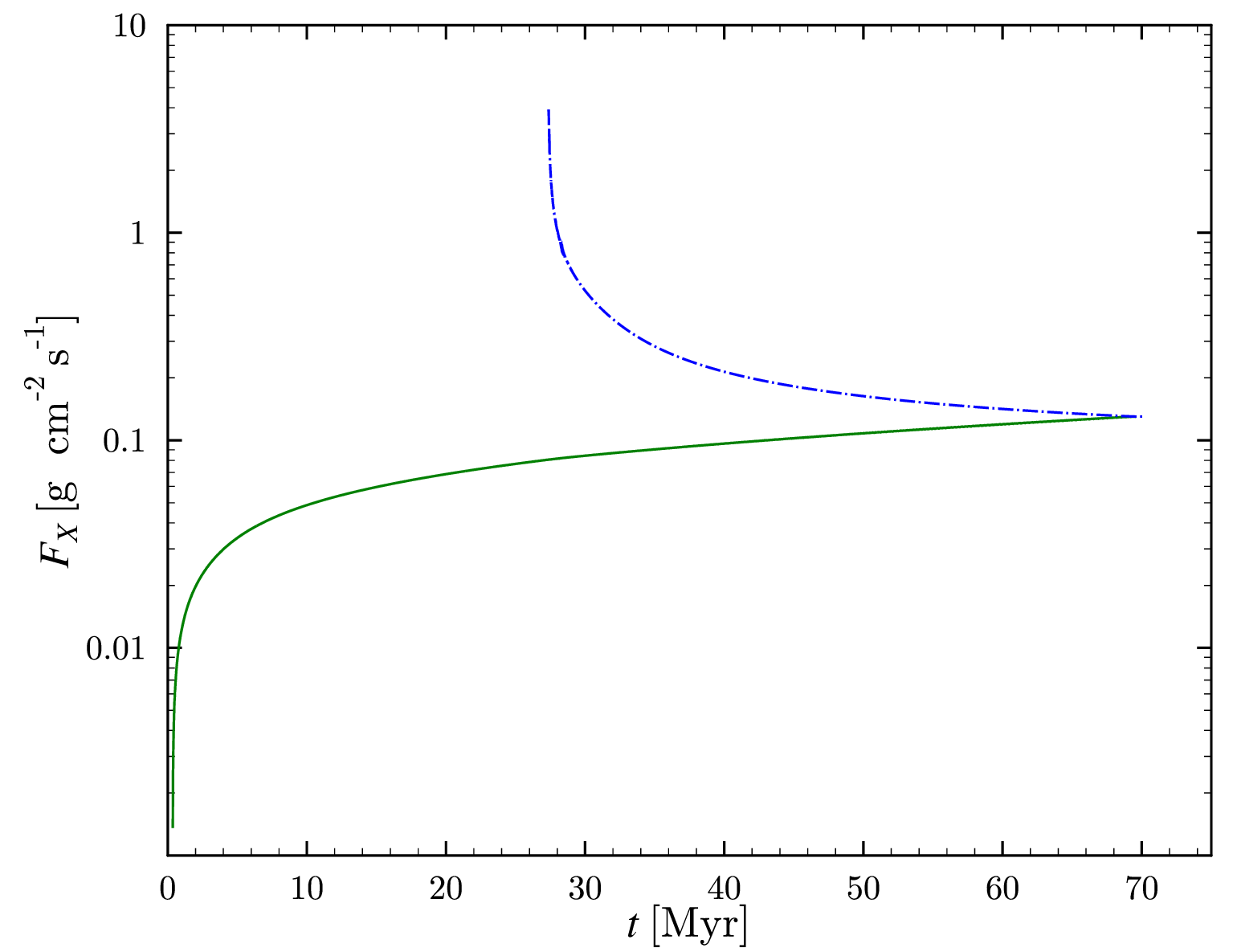}
\end{center}
\vspace{-0.2cm}
\caption[ ]{Light element flux vs. time.
Dot-dashed (blue) and solid (green) curves show fluxes based on the fully 
crystallized core growth rate and on the gradual release of light 
elements controlled by the latent heat removal, respectively.} 
\label{flux}
%\end{figure*}
\end{figure}

We have calculated the flux of light elements using both 
prescriptions and show the results in Fig.\ \ref{flux}. Based on the 
phase diagram of \citet{BD21}, we have set the difference in the carbon mass 
fraction between the liquid and the solid to $\Delta X = 0.11$. 
Dot-dashed (blue) curve assumes that the flux is determined by the fully 
crystallized core growth rate. In agreement with the analysis of \citet{FCC24}, 
this curve is divergent. Solid (green) curve\footnote{To reduce 
discretization artefacts, we applied moving average over $\pm 1\%$ of the 
current time $t$.} takes into account gradual 
release of light elements controlled by the latent heat removal which 
has been taking place over $\sim 27$ Myr before the start of the fully 
crystallized core formation. In this case, as soon as the central region 
reaches the crystallization temperature, the flux rapidly grows starting 
from zero. Then it is everywhere finite, slowly growing, and, even at 
10 Myr since the appearance of the fully solidified core, it is more 
than two times smaller than the previous estimate. At the end of 
our modeling, $r_{\rm c}$ catches up with $r_{\rm m}$, and both 
calculations of the light element flux coincide as they should.             

%%%%%%%%%%%%%%%%%%%%%%%%%%%%%%%%%%%%%%%%%%%%%%%%%%%%%%%%%%%%
%%%%%%%%%%%%%%%%%%%%%%%%%%%%%%%%%%%%%%%%%%%%%%%%%%%%%%%%%%%%

\section{Conclusion}
We have considered thermal evolution of the internal region of a 
$0.9 \, M_\odot$ C/O WD over $\sim 70$ Myr since the beginning of 
ion plasma crystallization in its center. We have observed formation 
of two spherical shells. The outer one surrounds the region, 
where crystallization has begun, whereas the inner one bounds the fully 
solidified core, which has exhausted its latent heat. The region between
the shells is partially liquid and partially solid. It gradually emits 
the latent heat of crystallization, which is carried away along the 
existing thermal gradients, and also it releases light elements in the 
process of element redistribution, accompanying the mixture 
solidification. The inner shell first appears $\sim 27$ Myr after the 
appearance of the outer shell, whose radius reaches $\sim 1000$ km at 
this moment. At $\sim 70$ Myr, the inner shell catches up with the outer 
one, and the crystallization process adopts the familiar form of a 
motion of a 2D front.   

Assuming that all released light elements cross the outer shell, we have
estimated their flux induced by the mixture crystallization. Up until 
the time when both shells merge, the result obtained is significantly 
smaller than the estimate based on the model of \citet{FCC24}. In particular, 
according to Fig. 2 of that work, the efficient convection driven by the 
light element flux is expected to cease after $\sim 3$ Myr since the 
formation of the fully solidified core. However, at this moment, the 
actual flux of light elements is $\sim 5$ times lower than the \citet{FCC24} 
prediction and is even lower at earlier times (cf. Fig.\ \ref{flux}). 
This means that the criterion for the efficient convection is not 
fulfilled at the early stages of WD crystallization and, consequently, 
such a phenomenon likely does not occur.

\begin{acknowledgements}

%\section*{Acknowledgments}
%
The author is deeply grateful to D.\ G.\ Yakovlev for discussions and to 
expert referee, Mike Montgomery, for valuable comments. This work 
has been supported by the Russian Science Foundation grant 24-12-00320. 

\end{acknowledgements}

%\section*{Data Availability}
%%
%The data underlying this article will be shared on reasonable 
%request to the author.

\bibliographystyle{aa}

\end{document}